\documentclass[conference]{IEEEtran}
\IEEEoverridecommandlockouts
\usepackage{cite}
\usepackage{amsmath,amssymb,amsfonts}
\usepackage{algorithmic}
\usepackage{graphicx}
\usepackage{textcomp}
\usepackage{hyperref}
\usepackage{xcolor}
\usepackage{url}
\usepackage{float}
\usepackage{multirow}
\usepackage[most]{tcolorbox}
\usepackage{listings}
\usepackage{array}

\lstdefinestyle{SQL}{
  language=SQL,
  basicstyle=\small\ttfamily,
  keywordstyle=\color{blue}\bfseries,
  commentstyle=\color{gray},
  stringstyle=\color{orange},
  numbers=left,
  numberstyle=\tiny,
  stepnumber=1,
  numbersep=5pt,
  backgroundcolor=\color{white},
  frame=lines,
  tabsize=4,
  showspaces=false,
  showstringspaces=false
}

\tcbuselibrary{listings,skins}
\tcbset{
  SQLbox/.style={
    enhanced,
    colback=white,
    colframe=blue!20!black,
    sharp corners,
    frame hidden,
    interior hidden,
    boxsep=0pt,
    top=0pt,
    bottom=0pt,
  }
}

\newtheorem{Example}{Example}
\newcommand{\LineageX}{\textsc{LineageX}}
\definecolor{lstpurple}{rgb}{0.5,0,0.5}

\lstdefinestyle{psqlcolor}
{
tabsize=2,
basicstyle=\footnotesize\upshape\ttfamily,
language=SQL,
extendedchars=false,
keywordstyle=\bfseries\color{lstpurple},
deletekeywords={count,min,max,avg,sum},
keywordstyle=[2]\color{lstblue},
stringstyle=\color{lstreddark},
commentstyle=\color{lstgreen},
mathescape=true,
escapechar=@,
sensitive=true,
numbers=left
}

\def\BibTeX{{\rm B\kern-.05em{\sc i\kern-.025em b}\kern-.08em
    T\kern-.1667em\lower.7ex\hbox{E}\kern-.125emX}}
\begin{document}

\title{\LineageX{}: A Column Lineage Extraction System for SQL\\}

\author{\IEEEauthorblockN{Shi Heng Zhang}
\IEEEauthorblockA{%\textit{Computing Science} \\
\textit{Simon Fraser University}\\
Burnaby, Canada \\
andy\textunderscore zhang@sfu.ca}
\and
\IEEEauthorblockN{Zhengjie Miao}
\IEEEauthorblockA{%\textit{Computing Science} \\
\textit{Simon Fraser University}\\
Burnaby, Canada \\
zhengjie@sfu.ca}
\and
\IEEEauthorblockN{Jiannan Wang}
\IEEEauthorblockA{%\textit{Computing Science} \\
\textit{Simon Fraser University}\\
Burnaby, Canada \\
jnwang@sfu.ca}
}

\maketitle

\begin{abstract}
As enterprise data grows in size and complexity, column-level data lineage, which records the creation, transformation, and reference of each column in the warehouse, has been the key to effective data governance that assists tasks like data quality monitoring,  storage refactoring, and workflow migration. Unfortunately, existing systems introduce overheads by integration with query execution or fail to achieve satisfying accuracy for column lineage. In this paper, we demonstrate \LineageX{}, a lightweight Python library that infers column-level lineage from SQL queries and visualizes it through an interactive interface. 
\LineageX{} achieves high coverage and accuracy for column lineage extraction by intelligently traversing query parse trees and handling ambiguities.
The demonstration walks through use cases of building lineage graphs and troubleshooting data quality issues. \LineageX{} is open sourced at \url{https://github.com/sfu-db/lineagex} and our video demonstration is at \url{https://youtu.be/5LaBBDDitlw}
\end{abstract}

\begin{IEEEkeywords}
database, lineage, provenance
\end{IEEEkeywords}

\vspace{-5pt}
\section{Introduction} \label{intro}
% Lineage, or provenance in its typical form, depicts how data is gathered, identifies where data originates, and traces the changes in data throughout its development~\cite{buneman2001and,cui2003lineage}. In a data warehouse setting, lineage information can be used to track how data flows from upstream artifacts to the downstream services, e.g., how a change in the upstream would affect the downstream. As the complexity of the schema grows in the data warehouse, finer-grained \emph{column-level lineage} is often used for reducing the tracing space from source columns to downstream columns. 
%detect changes in the upstream and how that change would affect its downstream services or databases. 
% Another real-world scenario of the lineage information would be evaluating the trustworthiness of the data~\cite{trust_prov} for improving the overall data quality and checking data compliance with regulations, such as GDPR and HIPAA. In this case, column-level lineage can help identify how the sensitive data flows in the entire pipeline.
Data governance has become increasingly crucial as data is becoming larger and more complex in enterprise data warehouses. For example, in an organization's data pipeline, data flows from upstream artifacts to downstream services, which may be built by various teams that know little about other teams' work and often introduce challenges when anyone wants to change their data.
In this case, lineage~\cite{buneman2001and,cui2003lineage}, especially finer-grained \emph{column-level lineage}, is often needed for simplifying the impact analysis of such a change, i.e., how a change in the upstream would affect the downstream. 
In another real-world scenario, column-level lineage can help identify how sensitive data flows throughout the entire pipeline, thereby improving the overall data quality and validating data compliance with regulations, such as GDPR and HIPAA~\cite{trust_prov}.

% While many existing systems like Atlas~\cite{Atlas} and DataHub~\cite{DataHub} support in-DBMS lineage tracing, there remains the need to curate the lineage information from static analysis of queries or query logs. 
While capturing lineage information in DBMS has been studied extensively in the database community~\cite{perm,provsql,gprom}, the need remains to curate the lineage information from static analysis of queries (without executing the queries).
On the one hand, existing systems or tools would introduce large overheads by either modifying the database internals~\cite{perm,provsql} or rewriting the queries to store the lineage information~\cite{gprom,hernandez2021computing}. On the other hand, different data warehouse users may need to disaggregate the lineage extraction workflow from query execution to simplify their collaboration, as shown in the following example.

\begin{figure}
  \centering
  \includegraphics[width=\linewidth]{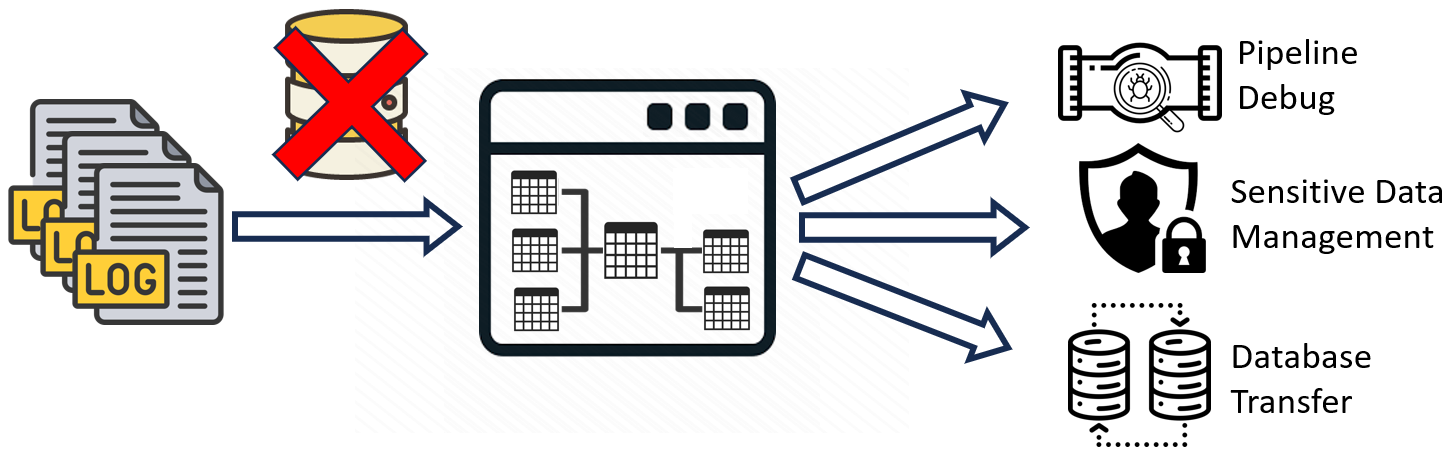}
    \vspace{-24pt} % Adjust the value to control the amount of space
  \caption{Lineage extraction from query logs without a database connection.}
  \label{fig:intro}
    \vspace{-18pt} % Adjust the value to control the amount of space
\end{figure}

\begin{Example} \label{sec:example}
% \vspace{-5pt}
% An online shop uses a data warehouse to store and analyze its customer and transaction data. There is a view, \texttt{webinfo}, which keeps track of user activities, and another view, \texttt{info}, connects the users' website activities (stored in view \texttt{webact}) to their orders, which may be used for recommendation purposes. However, the online shop owner found that the view \texttt{info} has stopped updating since the start of 2023 and asked for help from the data warehouse provider.
An online shop uses a data warehouse to store and analyze its customer and transaction data. There is a view, \texttt{webinfo}, which keeps track of user activities, and another view, \texttt{info}, connects the users' website activities (stored in view \texttt{webact}) to their orders, which may be used for recommendation purposes. However, the online shop owner decides to edit the \texttt{page} column of the \texttt{web} table and requests an impact analysis from the data warehouse provider.

%so what is the fastest way to identify the root cause?
\vspace{-10pt}
\begin{tcolorbox}[SQLbox]
\begin{lstlisting}[style=psqlcolor,mathescape,deletendkeywords={YEAR, DATE}]
$Q_1=$ CREATE VIEW info AS
SELECT c.name, c.age, o.oid, w.*
FROM customers c JOIN orders o ON c.cid = o.cid
JOIN webact w ON c.cid = w.wcid;
$Q_2=$ CREATE VIEW webact AS
SELECT w.wcid, w.wdate, w.wpage, w.wreg 
FROM webinfo w 
INTERSECT
SELECT w1.cid, w1.date, w1.page, w1.reg
FROM web w1;
$Q_3=$ CREATE VIEW webinfo AS
SELECT c.cid AS wcid, w.date AS wdate, 
  w.page AS wpage, w.reg AS wreg
FROM customers c JOIN web w ON c.cid = w.cid
WHERE EXTRACT(YEAR from w.date) = 2022;
\end{lstlisting}
\vspace{-10pt}
\end{tcolorbox}

%The engineer assigned to the task is not part of the team that built the data pipeline. 
Due to access control and privacy regulations, the engineer from the data warehouse provider can only access the log of database queries instead of the DBMS. The task is prone to being time-consuming and may involve tracing unnecessary columns without a comprehensive data flow overview. To address this, the engineer considers using tools like SQLLineage~\cite{sqllineage} to extract and visualize the lineage graph.

%and understand dependencies and relationships among tables and columns. %Centering the lineage graph around the target table filters out irrelevant columns, offering a clear starting point for investigation.
Although it can generate a lineage graph as shown in \autoref{fig:example}, there are a few issues with the column lineage. One is that the node of \texttt{webact} erroneously includes four extra columns, highlighted in a solid red rectangle. 
%between views \texttt{webinfo} and \texttt{webact}. %due to a set operation. Consequently, the node of \texttt{webact} erroneously includes four extra columns 
Another error arises for view \texttt{info} due to the \texttt{SELECT *} operation, which makes it unable to match the output columns to columns in \texttt{webact}. Instead, it would return an erroneous entry of \texttt{webact.*} to \texttt{info.*} (in solid red rectangle) while omitting the four correct columns from \texttt{webact}. It would also return fewer columns for the view \texttt{info} (in dashed red rectangle) and completely ignore the edges connecting \texttt{webact} to it (the yellow dashed arrows). If the engineer used the information from this lineage graph, then not only an erroneous column (\texttt{webact.page}) is provided, but the results also miss actual impacted columns from the \texttt{webact} and \texttt{info} table. 
% This issue originates from the existence of \texttt{SELECT w.*}, which leads to ambiguities and inaccuracies in the lineage graph. 
As we will demonstrate, our approach is able to handle statements like \texttt{SELECT w.*} and capture all columns and their dependencies missed by prior tools.
 \begin{figure}
  \centering   
  \includegraphics[width=0.4\textwidth, height=4cm]{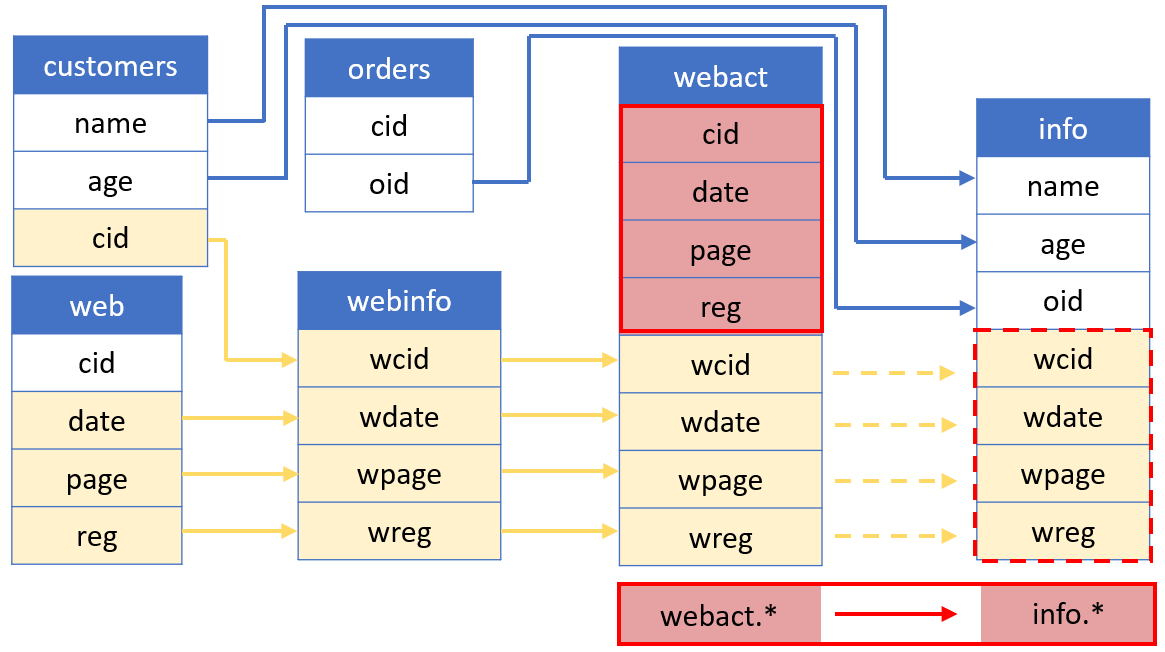}
  \vspace{-10pt} % Adjust the value to control the amount of space
  \caption{The lineage graph for Example~\ref{sec:example}. Existing tools like SQLLineage~\cite{sqllineage} would miss columns in the dashed red rectangle and return wrong entries in the solid red rectangle, while the yellow is the correct lineage}
  \label{fig:example}
  \vspace{-15pt}
\end{figure}

\end{Example}

Curating lineage information from query logs is also advantageous for debugging data quality issues, enhancing data governance, refactoring data, and providing impact analysis. However, existing tools~\cite{sqlglot,sqllineage} often fail to accurately infer column lineage due to the absence of metadata. To support developers and analysts in extracting lineage without the overhead of running queries in DBMS, 
we develop a lightweight Python library, \LineageX{}, which \emph{constructs a column-level lineage graph from the set of query definitions and provides concise visualizations of how data flows in the DBMS. }

% To have a lineage graph with great accuracy, we present LineageX\footnote{https://github.com/sfu-db/lineagex}, a lightweight Python library tailored for developers and analysts. From the provided SQLs, LineageX adeptly uncovers lineage information within SQL queries and base tables, offering a concise visualization with user interactions. With these capabilities, LineageX encounters two technical challenges: i) efficiently and accurately inferring column lineage, and ii) addressing metadata inference in the absence of a database connection.

\noindent\textbf{Challenges.} \LineageX{} achieves accurate column-level lineage extraction by addressing the following two challenges.
First is the \emph{\underline{variety of SQL features}}, especially for features that involve intermediate results or introduce column ambiguity. For example, Common Table Expressions (CTEs) and subqueries generate intermediate results that the output columns depend on, while the desired lineage should only reveal the source tables and columns. Set operations may introduce column ambiguity, primarily due to the lack of table prefixes.
% \textcolor{red}{@Andy: 1. Some detail for the column ambiguity; 2. Is it possible to add this to the example?}
Second, when there is an \emph{\underline{absence of metadata}} from the DBMS on each table's columns, e.g., when the query uses \texttt{SELECT *} or refers to a column without its table prefix, it may introduce ambiguities. Thus, prior works fail to trace the output columns when the \texttt{*} symbol exists and cannot identify their original table without an explicit table prefix.

\noindent\textbf{Our contributions.} For the first challenge, \LineageX{} uses a SQL parser to obtain the queries' abstract syntax trees (AST) and perform an \emph{intelligently designed traversal on the AST with a comprehensive set of rules to identify column dependencies}. \LineageX{} addresses the second challenge by \emph{dynamically adjusting the processing order for queries} when it identifies ambiguities in the source of tables or columns. 
% The solutions are simple yet effective and have been proven more accurate than prior tools.
Moreover, to accommodate the majority of data practitioners, we integrate \LineageX{} with the popular Python data science ecosystem by providing a \emph{simple API} that directly takes the SQL statements and outputs the lineage graph. Besides the API, we provide a UI that \emph{visualizes the column lineage} for users to examine.

% Another consideration is integration with Python. According to an annual survey by JetBrains, SQL consistently ranks in the top 3 languages used alongside Python, with over 60\% of answers indicating they work with a relational database~\footnote{Python Developers Survey, https://lp.jetbrains.com/python-developers-survey-2022}.
% Therefore, we decide to integrate our tool with the popular Python data science ecosystem to accommodate the majority of data practitioners.

% In this demonstration, we will showcase pipeline debugging scenarios and illustrate how \LineageX{} provides accurate column-level lineage and can further help the user monitor their data flow. The user will be able to compare the lineage extraction results by \LineageX{} with prior tools, including SQLLineage~\cite{sqllineage} and SQLGlot~\cite{sqlglot}.
% Since pre-trained large language models (LLMs) have shown impressive performance in understanding code, including SQL queries~\cite{nam2024using,gao2023text}, we also include the baseline of using GPT-3.5 for column lineage extraction in the demo.

In this demonstration, we will showcase the impact analysis scenario and illustrate how \LineageX{} provides accurate column-level lineage to further help users monitor their data flow. The user can compare the lineage extraction results by \LineageX{} with prior tools. Since pre-trained large language models (LLMs) have shown impressive performance in understanding code,
we will also demonstrate using state-of-the-art LLMs like GPT-4o for impact analysis and how to augment their results with the column-level lineage from \LineageX{}.
% including SQL queries~\cite{nam2024using,gao2023text}, its robustness will also be mentioned.

% From the survey, a majority of the answers do work with a relational database, so another consideration is the ability to work alongside a database instance. A database instance could provide more information such as metadata and physical plan for the SQLs. LineageX is able to produce the lineage graph between the views and tables in the database or input SQLs in a file. 

\begin{figure}
  \centering
  \includegraphics[width=0.92\linewidth]{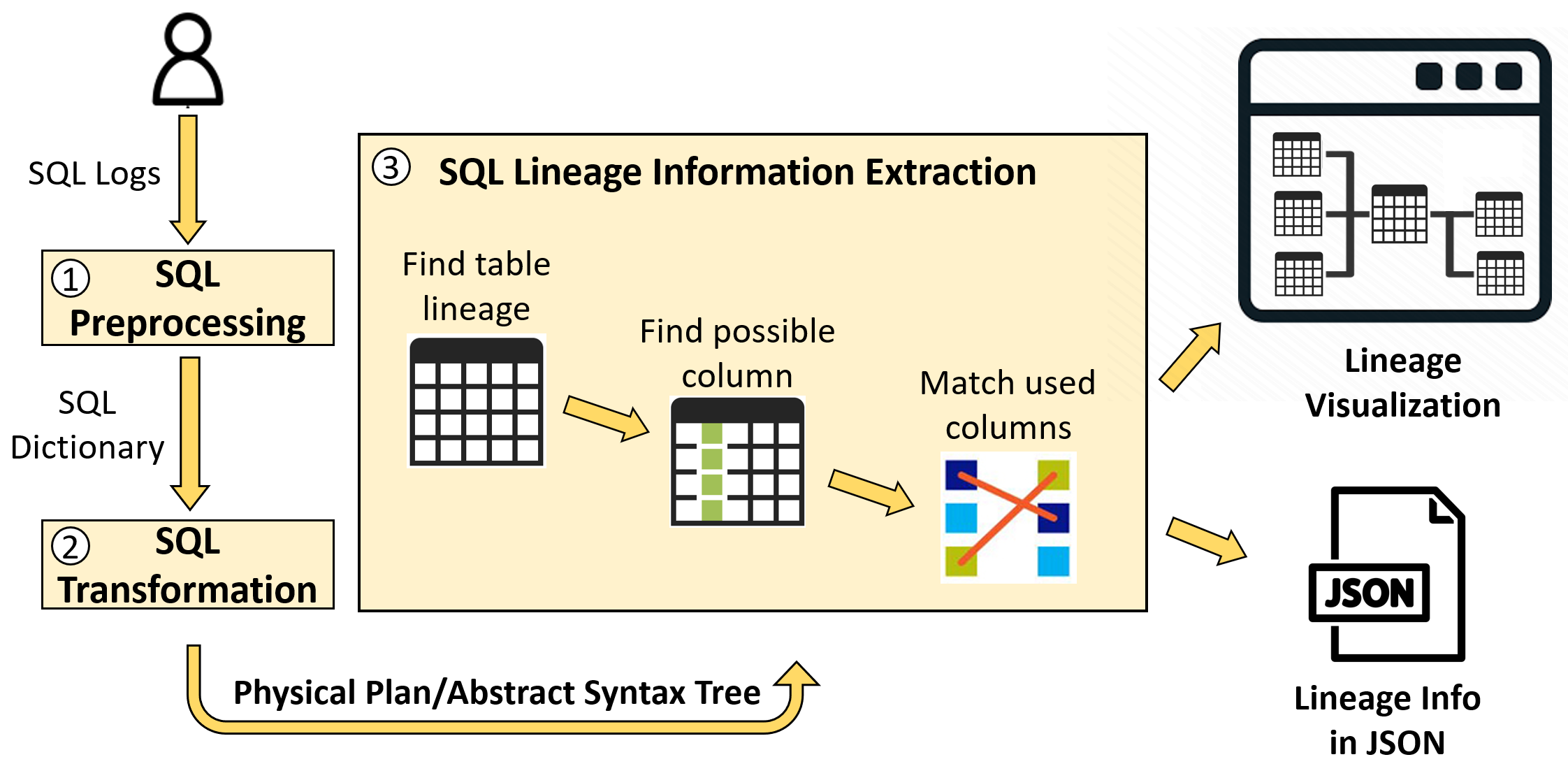}
  \vspace{-12pt}
  \caption{An illustration of \LineageX{}.}
  \label{fig:pipeline}
  \vspace{-20pt} % Adjust the value to control the amount of space
\end{figure}

% \section{RELATED WORK \& BACKGROUND}
\section{Background and Related work}

Data lineage tracks how data flows between each step in a data processing pipeline. Consider each processing step as a query $Q$, the \emph{table-level lineage} $\mathtt{T}$ of $Q$ encodes which input tables contribute to its output;  and the \emph{column-level lineage} $\mathtt{C}$ is a mapping from $Q$’s output columns $\mathcal{C}^{output}$ to $Q$’s input columns ${C}^{source}$, which encodes for each output column which specific columns in the input tables it relies on. More specifically, for an output column $c^{out} \in \mathcal{C}^{output}$ of $Q$, an input column $C^{src} \in \mathcal{C}^{source}$ is included in $\mathtt{C}(c^{out})$ if any changes to $C^{src}$ will lead to a potential change in the values in $c^{out}$ — we may not only include the input columns directly contributing to the output value but also take any column referred in the query into consideration.

Then, consider a set of queries $\mathcal{Q} = \{Q_i\}$, \emph{lineage extraction} is to find the pair $(\mathtt{T}_i, \mathtt{C}_i)$ for each $Q_i$. Note that queries in $\mathcal{Q}$ may be table/view creation queries, hence $\mathtt{T}_i$ and $\mathtt{C}_i$ may map the outputs of $Q_i$ to the outputs of other queries. In practice, to make the lineage graph easy to read, we can combine these two graphs and group all columns' output by the same query to visualize this graph.

\smallskip\noindent\emph{Related work.}
Data lineage~\cite{buneman2001and,cui2003lineage} has been studied extensively in the database research community. To track fine-grained lineage information down to the tuple level or cell level, people have extended relational database engines like in ProvSQL~\cite{provsql} and PERM~\cite{perm} or built middlewares that rewrite queries~\cite{gprom,hernandez2021computing}, which are often "overkill" for column-level lineage.
Various industry-leading tools, including Linkedin's Datahub~\cite{data_versioning}, Microsoft's Purview~\cite{purview}, and Apache Atlas~\cite{Atlas}, are more than capable of handling data pipelines and relational databases, but they may incur high operational and maintenance costs. % and consume additional computing resources. 
%, and may expose drawbacks, including challenges in supporting various databases and the need for manual entries. 
Vamsa~\cite{namaki2020vamsa} annotates columns used to train machine learning models for Python scripts.
There are also Python libraries like SQLGlot~\cite{sqlglot} and SQLLineage~\cite{sqllineage} that parse SQL queries statically; however, they focus on lineage for individual files, lacking the ability to find the dependency across queries, especially when there are ambiguities in table or column names. % due to the \texttt{*} symbol and missing table prefixes. 
%These tools are integrated into relational database engines, requiring to execute the queries and thus may lead to large performance and storage overheads. 
None of the methods above provides lightweight and accurate lineage extraction at the column level, like what \LineageX{} offers, without running the database queries; \LineageX{} can also visualize related tables and the data flow between columns in an interactive graph.

\begin{table}[t]
\caption{Keyword Rules.}
\label{rule_table}
\vspace{-7pt} % Adjust the value to control the amount of space
{\scriptsize
\begin{tabular}{|>{\centering\arraybackslash}p{0.9cm}|>{\centering\arraybackslash}p{3.9cm}|>{\centering\arraybackslash}p{2.7cm}|}

\hline
\textbf{Keyword}                                                          & \textbf{Process}                                                                                                                                                                                                                     & \textbf{Explanation}                                                                                                                                                \\ \hline
\textbf{SELECT}                                                           & \textit{$C^{con} \gets p \cup C^{pos}\forall p \in \mathcal{P}$}                                                                                                                                                  & \begin{tabular}[c]{@{}c@{}}Resolve $C^{con}$\\ for each projection\end{tabular}                                                                        \\ \hline
\textbf{\begin{tabular}[c]{@{}c@{}}FROM\\ (Table/\\ View)\end{tabular}}   & \textit{\begin{tabular}[c]{@{}c@{}}$T \gets T \cup $ \{this table\}\\ $C^{pos} \gets C^{pos} \cup $ \{its columns\}\end{tabular}}                                                                          & \begin{tabular}[c]{@{}c@{}}Add to $T$ for table lineage,\\ and add its column to $C^{pos}$ %\\ for column candidates
\end{tabular}                       \\ \hline
\textbf{\begin{tabular}[c]{@{}c@{}}FROM\\ (CTE/\\ Subquery)\end{tabular}} & \textit{\begin{tabular}[c]{@{}c@{}}find the CTE/Subquery in $M_{CTE}$\\ $C^{pos} \gets C^{pos} \cup $ \{its columns\}\end{tabular}}                                                                              & \begin{tabular}[c]{@{}c@{}}Find this CTE/subquery \\ in $M_{CTE}$, and add its \\ columns to $C^{pos}$\end{tabular}                                          \\ \hline
\textbf{\begin{tabular}[c]{@{}c@{}}WITH/\\ Subquery\end{tabular}}         & \textit{\begin{tabular}[c]{@{}c@{}}$M_{CTE} \gets T, C^{con}, C^{ref}$\\ $T, C^{con}, C^{ref},$ $C^{pos}, \mathcal{P} \gets \emptyset$\end{tabular}}                             & \begin{tabular}[c]{@{}c@{}}$M_{CTE}$ gets all the current \\ table and column lineage, \\ store to be referenced\end{tabular}                                             \\ \hline
\textbf{\begin{tabular}[c]{@{}c@{}}Set\\ Operation\end{tabular}}          & \textit{\begin{tabular}[c]{@{}c@{}}$C^{ref} \gets C^{ref} \cup p \cup C^{pos} $ $\forall p \in \mathcal{P}$\\ $C^{pos}, \mathcal{P} \gets \emptyset$\\ repeat for other leaves\end{tabular}} & \begin{tabular}[c]{@{}c@{}}Add all the columns in the\\ projection to $C^{ref}$,\\ this process is repeated for \\ other leaves connected\end{tabular} \\ \hline
\textbf{\begin{tabular}[c]{@{}c@{}}Other \\ Keywords\end{tabular}}        & \textit{\begin{tabular}[c]{@{}c@{}}$temp\textsubscript{cols} \gets$ all columns here\\ $C^{ref} \gets C^{ref} \cup p \cup C^{pos}$\\ $\forall p \in temp\textsubscript{cols}$\end{tabular}}   & \begin{tabular}[c]{@{}c@{}}Add all the columns\\ found here to $C^{ref}$\end{tabular}                                                                 \\ \hline
\end{tabular}
}
%\vspace{-20pt} % Adjust the value to control the amount of space
\vspace{-12pt}
\end{table}

\section{SYSTEM and IMPLEMENTATION} \label{detail}
The overview of the \LineageX{} system is shown in \autoref{fig:pipeline}. \LineageX{} allows users to input a list of SQL statements or query logs.
% Then, the queries are preprocessed to collect the metadata for each table/view, transformed into a tree structure, and finally, the \emph{lineage information extraction} module infers the column lineage by traversing the query trees. 
Below are details of each module.

\smallskip\noindent\emph{SQL Preprocessing Module.} The first step is to scan each query and record the mappings from the query's identifier to its query body. 
% For each query, the Preprocessing Module removes any comments, extra spaces, and keywords before the first \texttt{SELECT}.
For \texttt{CREATE} statements, we use the created table/view's name as the query identifier, while for \texttt{SELECT} statement-only queries, we use a randomly generated id\footnote{For some systems like dbt, queries containing only \texttt{SELECT} statement are stored in separate files. In this case, we will use the file name as the query identifier. We also provide a dbt-specific wrapper for \LineageX{}.}.
Then, each identifier is mapped to the body of the \texttt{SELECT} statement, forming a key-value pair.
For instance, for $Q_3$ in Example~\autoref{sec:example}, our module would have \texttt{webinfo} as the key and the \texttt{SELECT} statement \ldots(line 12 to 15) as the value. These key-value pairs are stored in a Query Dictionary (QD), which will be further used to facilitate the inference between queries and identify the query dependencies.
 
 % Each parsed statement is assigned a representative name, forming a key-value pair in this module's output, the \textbf{SQL Dictionary}. The extracted name serves as the key, and the SQL statement serves as the value. 

\smallskip\noindent\emph{SQL Transformation Module.} Then, the Transformation Module reads each entry in the dictionary QD from the Preprocessing Module, generating an abstract syntax tree (AST) using a SQL parser (in the implementation, we used SQLGlot). The SQL AST captures all keywords and expressions in the query in a tree-like format, where the leaf nodes represent the initial scanning of source tables or the parameters of each operator, the root represents the final step, and intermediate nodes represent relational operators in the query.

\smallskip\noindent\emph{SQL Lineage Information Extraction Module.}
The final module takes each query AST as input and builds the mappings from the result view/table to its lineage \texttt{T} and the mapping from output columns $C^{output}$ to input columns $C^{source}$. We consider three types of columns in the lineage: 1)$C^{con}$: columns that directly contribute to $C^{output}$; 2) $C^{ref}$: columns referenced in the query, e.g., columns used in the join predicate or the \texttt{WHERE} clause; and %3) columns that directly contribute to $C^{output}$ and are also referenced in the query, denoted as $C^{both}$.
3) $C^{both}$: columns in both $C^{con}$ and $C^{ref}$.
The extraction process involves traversing the AST with a post-order Depth-First Search (DFS), for which we create some temporary variables:
$M_{CTE}$ is a mapping for the table and column lineage information from \texttt{WITH/subquery}, 
$C^{pos}$ denotes column candidates, and $\mathcal{P}$ denotes the resulting columns of the most recent projection. 
When encountering different keywords, the lineage information and temporary variables will be updated according to the rules in \autoref{rule_table}.

% From the problem definition, it states that \texttt{T} and \texttt{C} are part of the final output, where \texttt{T} denotes table lineage, and \texttt{C} denotes column-level lineage. There are also three types of dependency columns in the mapping of $C^{output} to $C^{source}: $C^{con}(columns that directly contribute to $C^{output}), $C^{ref}referenced columns, and $C^{both}(columns with both these properties). Some temporary variables will be utilized during the extraction: $M_{CTE}$ is a mapping containing the table and column lineage information from \texttt{WITH/subquery}, $C^{pos} to denote column candidates for deducing column lineage, and \texttt{proj} to denote the most recent projection. The system employs a post-order Depth-First Search (DFS) traversal of the plan or the AST, with rules fit for different keywords, and the rules are detailed in \autoref{rule_table}.

\begin{figure}
  \centering
  \includegraphics[width=0.85\linewidth]{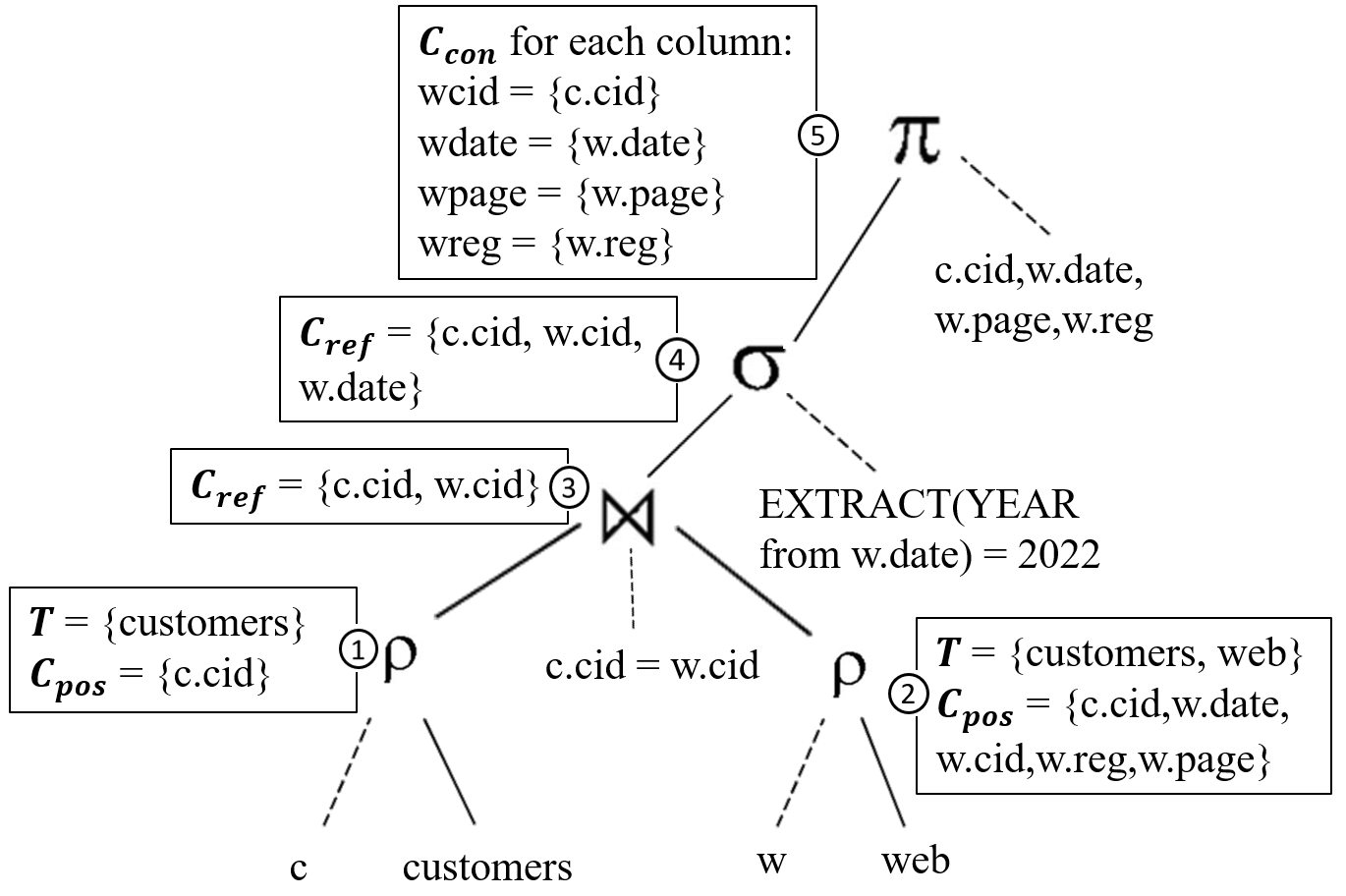}
  \vspace{-10pt} % Adjust the value to control the amount of space
  \caption{Sample AST and traverse order}
  \label{fig:algo}
  \vspace{-15pt} % Adjust the value to control the amount of space
\end{figure}

\begin{figure*}[ht]
  \centering
  \includegraphics[width=0.90\textwidth]{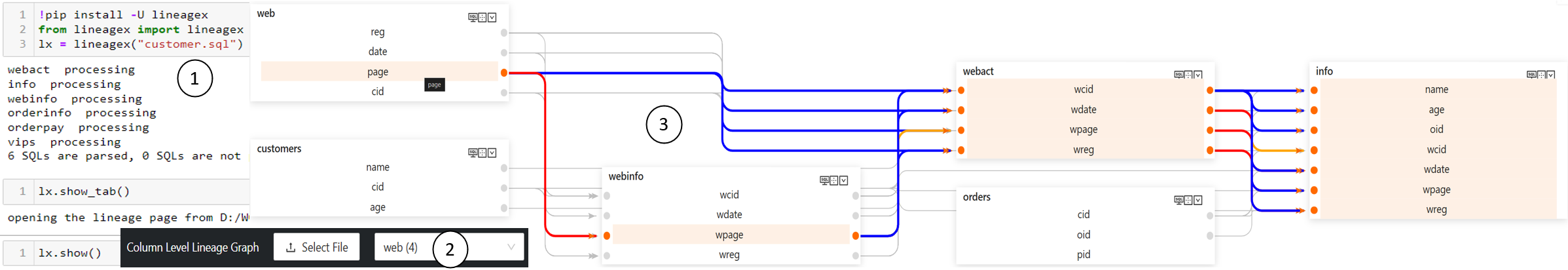}
  \vspace{-10pt} % Adjust the value to control the amount of space
  \caption{The User Interface of \LineageX{}.}
  \label{fig:demo_pic}
  \vspace{-15pt} % Adjust the value to control the amount of space
\end{figure*}

% The general traversal steps involve finding the lineage table, deducing possible output columns from those tables, matching them with columns referred to in the SQL, and producing the column lineage. 
An example for traversing the AST of $Q_3$ is shown in \autoref{fig:algo}. \textcircled{\raisebox{-0.9pt}{1}}: The traversal starts with the leaf node, scanning of \texttt{customers}, so it follows the \texttt{FROM} Rule by adding it to \texttt{T} and its columns to $C^{pos}$.
\textcircled{\raisebox{-0.9pt}{2}}: The next node is scanning of \texttt{web}, so it is added to \texttt{T} and its columns to $C^{pos}$.
\textcircled{\raisebox{-0.9pt}{3}}: The next node is a \texttt{JOIN}, following the \texttt{Other keywords} Rule: \texttt{customers.cid} and \texttt{web.cid} are added to to $C^{ref}$.
\textcircled{\raisebox{-0.9pt}{4}}: For the \texttt{WHERE} node ($\sigma$), same rule applies, hence adding \texttt{web.date} to $C^{ref}$. 
\textcircled{\raisebox{-0.9pt}{5}}: The last node is the \texttt{SELECT} ($\pi$), applying the \texttt{SELECT} Rule. Each output column's $C^{con}$ only has one column, e.g., \texttt{wcid} has $C^{con}$ of \texttt{customers.cid}.%, and column can be in both $C^{con}$ and $C^{ref}$.

% \subsection{Table/View Auto-Inference.} 
\smallskip\noindent\emph{Table/View Auto-Inference.} 
In the Lineage Information Extraction module, the system gives priority to SQL statements identified by keys in QD from the Preprocessing Module. This procedure leverages a stack to reorder the query ASTs to traverse, where current traversal is temporarily deferred and placed onto the stack. That is, in cases where the tables or views encountered during the traversal have not been processed yet, they are pushed to the stack. Once the lineage information of missing tables is extracted, the deferred operation is popped from the stack following a Last-In-First-Out protocol and resumes. This strategic approach plays a pivotal role in handling \texttt{SELECT *} statements and resolving ambiguities related to columns without a prefixed table name. 
\smallskip\noindent\textit{When the database connection is available.}
While primarily focusing on \emph{static} lineage extraction from query logs, \LineageX{} can also incorporate the extraction with a database connection. We extended \LineageX{} using PostgreSQL's \texttt{EXPLAIN} command to obtain the physical query plan instead of the AST from the parser, which provides accurate metadata to deal with table and column reference ambiguities. 
% The query plan offers several advantages, including the assurance of leveraging metadata to address \texttt{SELECT *} queries effectively. 
Similar to the 
absent views or tables in the static extraction, an error may occur due to missing dependencies when running the \texttt{EXPLAIN} command. This requires the stack mechanism and performing an additional step to create the views first to ensure the presence of the necessary dependencies.
\section{DEMONSTRATION} \label{demo}

% Lineage tools serve various purposes in data governance, data exploration, and the identification of data pipeline errors. In our demonstration, we employ the MIMIC dataset, which is hosted  here\footnote{https://zshandy.github.io/lineagex-demo/}. For a more comprehensive showcase, please refer to the environment detailed in Example \ref{sec:example}, where it will include additional SQL statements to simulate a realistic workload. Our primary objective remains discovering the root cause behind the halted updates for \texttt{info}.

We will walk through the audience with the use cases like Example~\ref{sec:example}, employing multiple datasets, such as the MIMIC dataset~\footnote{\url{https://github.com/MIT-LCP/mimic-code}} in the healthcare domain. The MIMIC dataset has a reasonably complex schema with more than 300 columns in 26 base tables and 700 columns in 70 view definitions. We demonstrate in detail each step of using \LineageX{} for our running example in the environment of a Jupyter Notebook.
% Our primary objective remains discovering the root cause behind the halted updates for \texttt{info}.

\textbf{Step 1: Get started.}  Users have the flexibility to store their SQL queries in either files or a Python list. In this example, all SQL queries are stored in the file \texttt{customer.sql}. Then the function call is straightforward, as outlined in Figure~\ref{fig:demo_pic} \textcircled{\raisebox{-0.5pt}{1}}, the users simply install and import the library, then call the \LineageX{} function. The result will be returned in a JSON file (lineage information) and an HTML file (lineage graph).

\textbf{Step 2: Locating the table.} Next, users can visualize the graph using the \texttt{show} function in the notebook or the \texttt{show\_tab} to open a webpage. Moreover, users can select the table of interest through a dropdown menu, as shown in Figure~\ref{fig:demo_pic} \textcircled{\raisebox{-0.5pt}{2}}. Subsequently, the target table \texttt{web} and its corresponding columns are displayed.

\textbf{Step 3: Navigating column dependency.} Users can click the \texttt{explore} button on the top right of the table to reveal the table's upstream and downstream tables, presenting the initial table lineage. The data flows from left to right on the visualization --- tables on the right are dependent on tables on the left. Since we are doing an impact analysis, that is to find all the downstream columns and their downstream columns and so on. The first \texttt{explore} action would only show \texttt{webinfo} and \texttt{webact} tables, since they are the only ones that are directly dependent on the \texttt{web} table. The next \texttt{explore} action would reveal the \texttt{info} table, and there would be no more downstreams for \texttt{info}. With the lineage graph, hovering over the \texttt{page} column highlights all of its downstream columns, as shown in Figure~\ref{fig:demo_pic} \textcircled{\raisebox{-0.5pt}{3}}.

\textbf{Step 4: Solving the case.} The \texttt{page} column directly contributes to \texttt{wpage} from \texttt{webinfo}(shown in red), so it is definitely impacted. The \texttt{webact} table is a result of a set operation from \texttt{web} and \texttt{webinfo}, therefore all of the \texttt{webact}'s columns will reference the \texttt{page} column and thus all get impacted(shown in blue and orange when it is both referenced and contributed). Since the \texttt{wcid} column is impacted, it is also in the \texttt{JOIN} operation for the \texttt{info} table, then all of the columns would reference the \texttt{wcid} column and potentially get impacted. Therefore, the end result for the impact analysis would be \texttt{webinfo.wpage} and all of the columns from the \texttt{webact} and \texttt{info} tables.

\smallskip\noindent\textbf{Comparison with existing methods.} In our demonstration, users can compare results from \LineageX{} with those from SQLLineage~\cite{sqllineage}. SQLLineage returns incorrect columns for \texttt{info} and lacks lineage information for columns derived from \texttt{webinfo}, as shown in \autoref{fig:example}. The users can also see how state-of-the-art LLMs respond to their questions about impact analysis: for example, GPT-4o is able to correctly identify all contributing columns impacted by changes to \texttt{page}—specifically, the \texttt{wpage} columns in \texttt{webinfo}, \texttt{webact}, and \texttt{info} tables (highlighted in red or orange), but it is not able to reveal the columns that are referenced (not directly contributing to) in the SQL (such as the \texttt{webact.wcid} in the \texttt{JOIN} condition).

% In more complex cases, like the MIMIC dataset, column lineage would help the LLM perform impact analysis accurately.
%Additionally, column lineage data enhances ChatGPT’s ability to assist with query optimization, database refactoring, and error analysis.

%Therefore, manual investigation is still much needed, visualization for an interaction-style analysis and the JSON format for a more comprehensive code-style analysis.

% \textbf{Solving the case.} As \texttt{wdate} only contains information from 2022, the user decides to trace its source. The user soon realizes that \texttt{webinfo.wdate} exhibits the same issue, while its source, \texttt{web.date}, updates normally. This observation leads to the conclusion that the issue likely lies with \texttt{webinfo}. Further investigation reveals that the problem is caused by the \texttt{WHERE} clause set to the year 2022. In comparison with another library, SQLLineage, its output presents incorrect columns for \texttt{info} and lacks lineage relationships for columns coming from \texttt{webinfo}, similar to what is depicted in \autoref{fig:example}.

% \noindent\textbf{Other scenarios.} Besides Jupyter notebook, users can also use \LineageX{} in a SQL-intensive environment like dbt~\cite{dbt}, where numerous SQL queries execute daily. We provide a dbt-specific wrapper for \LineageX{} \footnote{https://github.com/sfu-db/dbt-lineagex}, which works seamlessly as other dbt plugins.

% The process involves parsing the \texttt{manifest.json} for the SQLs, employing the same core functionality as LineageX 

\vspace{12pt}

\end{document}